\documentclass[11pt]{article}
\usepackage{hyperref}
\usepackage{amssymb}
\usepackage{graphicx}
\usepackage{slashed}
\usepackage{amsmath}
\usepackage{cite}
\usepackage{authblk}
\usepackage[section]{placeins}
\usepackage[a4paper, total={6.8in, 10in}]{geometry}
\numberwithin{equation}{section}
\title{Emergence of maximal acceleration from non-commutativity of space-time}
\author{E. Harikumar \thanks{harisp@uohyd.ernet.in}}
\author{Leela Ganesh Chandra Lakkaraju \thanks{Present Address: Harish-Chandra Research Institute, Chhatnag Road, Jhunsi, Allahabad-211019, Uttar Pradesh, India, e-mail:ganeshchandra@hri.res.in}}
\author{Vishnu Rajagopal \thanks{vishnurajagopal.anayath@gmail.com}}
\affil{School of Physics, University of Hyderabad, Central University P.O, Hyderabad-500046, Telangana, India}
\date{}
\begin{document}

\maketitle
\begin{abstract}
In this paper, we show that the causally connected $4$-dimensional line element of the $\kappa$-deformed Minkowski 
space-time induces an upper cut-off on the proper acceleration and derive this maximal acceleration, valid up to 
first order in the deformation parameter. We find a contribution to maximal acceleration which is independent of
$\hbar$ and thus signals effect of the non-commutativity alone. We also construct the $\kappa$-deformed geodesic equation and obtain its $\kappa$-deformed Newtonian limit, valid up to first order in deformation parameter. Using this, we constrain non-commutative parameters present in the expression for maximal acceleration. We analyse different limits of the 
maximal acceleration and also discuss its implication to maximal temperature. We also obtain a bound on the deformation 
parameter.
\\\\\textit{\textbf{Keywords : }}$\kappa$-space time, maximal acceleration, geodesic equation
\\\\\textbf{PACS Nos. : }11.10.Nx, 04.60.-m, 02.40.Gh, 04.50.Kd.
\end{abstract}

\section{Introduction}

One of the major stumbling blocks in our understanding of the universe is the lack of a consistent quantum theory of gravity. Irrespective of numerous approaches developed over the last several decades and substantial progress made, an entirely satisfactory microscopic theory of gravity still eludes us. Different frameworks to understand the quantum theory of gravity have pointed to certain characteristic features expected of this quantum gravity theory. Existence of a fundamental length scale, below which quantum gravity effects become important, is one of them \cite{review}.

The existence of a minimal length scale in some quantum gravity models has been shown to be associated with the emergence of an upper limit on the proper acceleration, known as maximal acceleration \cite{dk}. In \cite{cr2} it has been shown that covariant loop quantum gravity exhibits the existence of maximal acceleration, compatible with the local Lorentz symmetry. String theory has also predicted an upper bound on the string acceleration \cite{rp}. The notion of an upper bound on the value of allowed acceleration has a long history. The maximal acceleration in commutative space-time has been obtained in \cite{cai1}, from the 
$8$-dimensional phase space line element, by imposing the requirement that the events in the corresponding phase space to be time-like. Alternatively, the form of the maximal acceleration has been obtained in \cite{heis} using Heisenberg's uncertainty principle. This maximal acceleration has been used to smoothen the UV divergences in local QFT \cite{vvn}. Various aspects of kinematics and dynamics of a relativistic particle possessing maximal acceleration in $8$-dimensional space-time tangent bundle have been studied extensively \cite{scarp1}. Several authors have studied different phenomenological consequences of maximal acceleration models in \cite{cain}. The existence of a minimal time scale in physical phenomena leads to an upper cut-off on the allowed value of acceleration\cite{cald}, and using this argument, a model of an electron with maximal acceleration has been constructed and analysed\cite{cald1}. It has been shown that the maximal acceleration modifies the standard model of cosmology by introducing an inflationary expansion in the beginning stage \cite{gasp}.

Investigations of accelerating observers have resulted in unravelling curious results such as Unruh temperature\cite{unruh} and Hawking radiation\cite{hawking}. The implication of acceleration of the 
observer to the locality in general relativity had been studied in detail\cite{Mashhoon}. The study of physics of accelerated frames/observers has brought out a plethora of insights regarding different aspects of 
gravity. Thus it is of intrinsic interest to study possible limits placed on the acceleration of observers by any approach to quantum gravity, such as non-commutative geometry.

Models with maximal acceleration have been studied by various authors. In \cite{cai1} maximal acceleration was shown to arise as a consequence of geometrising quantum mechanics by introducing a metric in the 
8-dimensional phase-space. This was followed by analysing line element in the 8-dimensional phase space which depends not only on the co-ordinates $x^\mu$ but also on $E$ and ${\vec p}$, and using energy-momentum dispersion relation, a limit on the maximal acceleration of massive particle was derived. This maximal acceleration depends on the 
mass of the particle as well as on $\hbar$. Various aspects of this approach have been studied. In all these studies, the starting line element in 8-dimensions was not general covariant. In\cite{brandt}, this issue was solved by constructing a covariant line element by introducing non-linear connection in time-derivatives appearing in the velocity-dependent part of the 8-dimensional metric. In \cite{brandt}, the existence of maximal 
acceleration was argued and calculated by combining Unruh effect\cite{unruh} and Sakharov's result of the existence of a maximal temperature\cite{sakharov}. Another approach based on a pseudo-complex generalisation of Minkowski metric has been also shown to give a maximal cut-off value for acceleration. The line element used in this method \cite{schuller} was also covariant, like the one used in \cite{brandt}.

In all these approaches described above, one start with the assumption of form of a metric/ line element in 8-dimension space which has been taken as a 
direct sum of the metric of the Minkowski space-time and that of the corresponding tangent space. This 8-dimensional metric depends on the uniquely defined metric of the underlying Minkowski space-time. The basic entity of these calculations is 8-dimensional metric and hence the 
connection of the 4-dimensional Minkowski metric with experimental results is ambiguous. This shortcoming has been circumvented in an approach based on the effective theory of metrics \cite{torrome} and different geometric structures associated with maximal acceleration was studied in\cite{toller}. 

In the last decades, non-commutative geometry has been proposed as a paradigm to study the microscopic structure of 
space-time\cite{connes,fred}. One of the non-commutative space-times that has been attracting wide attention is 
the $\kappa$-deformed space-time\cite{luk, hopf}. This space-time appears in the low energy limit of certain loop 
gravity models. It is also the space-time associated with doubly(deformed) special relativity (DSR), which 
incorporate an additional constant having length dimension, apart from the velocity of light, in a consistent 
manner with the principle of relativity\cite{review,kow,majid}. Different aspects of this space-time have been 
investigated in recent times \cite{kreal, mel5, mel3, av1,av2}. Its implication to various phenomena were analysed 
in \cite{pach,mel6,trg,mel7,gauge,vishnu1,hari,kapoor,suman }. 
Recently, generalising the approach of\cite{cai1}, the maximal acceleration has 
been derived from the $8$-dimensional phase space metric defined in $\kappa$-deformed phase space \cite{vishnu4}. 
As in the commutative space-time, this study also used the 4-dimensional ($\kappa$-deformed) 
Minkowski metric to construct the 8-dimensional line element. We now initiate a construction which avoids 
the 8-dimensional metric to obtain the maximal acceleration in the $\kappa$-deformed space-time. Since the symmetry 
algebra associated with $\kappa$-deformed space-time has two fundamental constants, viz; $a$ having dimensions 
of the length and $c$, the velocity of light, it is possible to associate a natural time scale $\frac{a}{c}$ with 
phenomena in this space-time. This will have implications to the locality of accelerated observers\cite{Mashhoon} as 
well as on the allowed values of acceleration. Thus it is interesting to see whether one can get any limit on the 
acceleration within the $\kappa$-deformed Minkowski space-time, without resorting to the 8-dimensional space obtained 
by taking the direct sum of this space with its tangent space. In this paper, we address this issue and show that 
an upper cut-off on allowed value of acceleration naturally emerges from the 4-dimensional, $\kappa$-deformed line 
element and derive the expression of maximum acceleration valid up to first order in the deformation parameter $a$.
Though the $\kappa$-deformed metric we use also depends on the momentum $p_\mu$, unlike the earlier 
studies, this metric is not the direct sum of metric of 4-dimensional $\kappa$-Minkowski space-time and 
the metric of the corresponding momentum space. This expression of maximal acceleration derived here 
has two terms. One of them depends on the non-commutative parameters $a$ and $\alpha$, $\hbar$, the mass of the 
particle, and minimum distance of approach. The second term  is independent of $\hbar$, but depends 
on the non-commutative parameters $\alpha$, $\beta$, and the non-commutative energy scale, in addition to
the mass of the particle and a minimum distance of approach. This second term, 
which is independent of $\hbar$, emerges only due to non-commutativity of the space-time. 
This is a new feature of our result. We then 
derive the geodesic equation in the $\kappa$-deformed Minkowski space-time, valid up to first order in $a$ and 
obtain its Newtonian limit. This limit is seen to violate the equivalence principle and by comparing this with 
the experimental limit on the violation of equivalence principle, we obtain bounds on the 
dimensionless parameters($\alpha$ and $\beta$) appearing in the expression for maximal acceleration. We also analyse various limits of 
the maximal acceleration and discuss the implication to maximal temperature. 

The organisation of this paper is as follows. In sec.2, we construct the $\kappa$-deformed metric, valid up to 
the first order in the deformation parameter $a$, by using the generalised commutation relation between 
$\kappa$-deformed phase space co-ordinates. In sec.3, we show the emergence of the maximal acceleration in the 
$\kappa$-deformed space-time from the causally connected, $4$-dimensional, $\kappa$-deformed Minkowskian space-time. 
In sec.4, we derive the expression for the geodesic equation of a test particle in the $\kappa$-deformed 
space-time, valid up to first order in the deformation parameter, using the $\kappa$-deformed metric. In 
sub-sec.4.1, we obtain the $\kappa$-deformed Newtonian force equation from the $\kappa$-deformed geodesic 
equation and obtain a bound on the non-commutative parameters. In sec. 5, we summarise our results and give the 
concluding remarks. In appendix.A, we show that the consistency of the Newtonian limit of the deformed 
geodesic equation imply restriction on one of the dimensionless parameter ($\beta$) introduced through the
realisation of non-commutative co-ordinates. In appendix.B, we derive the first order corrections of maximal 
acceleration using the deformed uncertainity principle and show that, as expected, the result is different 
from that obtained from $4$-dimensional deformed line element (derived in Section 3). This signals the pure 
non-commutative effect captured by the maximal acceleration derived here from the $4$-dimensional line element 
defined on the $\kappa$-space-time. We use $\eta_{\mu\nu}=diag(1,-1,-1,-1)$. 

\section{$\kappa$-deformed metric}

In this section, we begin with a summary of the construction of metric in the $\kappa$-deformed 
space-time\cite{tajron}. The $\kappa$-deformed space-time co-ordinates can be realised in various ways; one can 
directly work with $\kappa$-deformed co-ordinates, or alternately one can represent the $\kappa$-deformed space-time 
co-ordinates as functions of commutative co-ordinate and their derivatives. We use the second method in our analysis. 
Using the generalised commutation relations between $\kappa$-deformed phase space co-ordinates, we show how to derive the 
$\kappa$-deformed metric, valid up to first order in $a$.

The $\kappa$-deformed space-time co-ordinates satisfy a Lie-algebra type commutation relations given by
\begin{equation}\label{nc}
[\hat{x}_{\mu},\hat{x}_{\nu}]=i(a_{\mu}\hat{x}_{\nu}-a_{\nu}\hat{x}_{\mu})
\end{equation}
where $a_{\mu}=(a,\overrightarrow{0})$. Here $a$ is the $\kappa$-deformation parameter having the dimension of $[L]$ and we choose a specific realisation for the $\kappa$-deformed space-time co-ordinate as \cite{tajron}
\begin{equation}\label{real}
\hat{x}_{\mu}=x_{\alpha}\varphi^{\alpha}_{\mu}(p).
\end{equation}
The consistency of Eq.(\ref{real}) and Eq.(\ref{nc}), up to order $a$, gives 
\begin{equation}\label{soln}
\varphi _{\mu}^{\alpha}(p)=\delta^{\alpha}_{\mu}\Big(1+\alpha\frac{a\cdot p}{\hslash}\Big)+\beta \frac{a^{\alpha}p_{\mu}}{\hslash}+\gamma \frac{p^{\alpha}a_{\mu}}{\hslash},
\end{equation}
where $\alpha,\beta,\gamma \in \mathbb{R}$ are dimensionless parmeters and they are constrainted by the relation $\gamma=\alpha +1$. Note that $\varphi_{\mu}^{\alpha}(p)$ 
reduces to $\delta_{\mu}^\alpha$ in the limit $a\to 0$ where as $\varphi_{\mu}^{\alpha}(p)$ diverges as $\hslash\to 0$. Hence we recover the commutative limit (where there is no $\hbar$ dependence) by taking $a\to 0$. The limit, $\frac{a}{\hslash}\to 0$ of $\varphi_{\mu}^\alpha(p)$ is also well defined and in this case also we recover the commutative co-ordinates. Thus using $\varphi^{\alpha}_{\mu}(p)$, Eq.(\ref{real}) is re-written as
\begin{equation}\label{x}
\hat{x}_{\mu}={x}_{\mu}\Big(1+\alpha\frac{a\cdot p}{\hslash}\Big)+
\beta \Big(x\cdot\frac{a}{\hslash}\Big)p_{\mu}+(\alpha+1)(x\cdot p)\frac{a_{\mu}}{\hslash}.
\end{equation}
Note that in the limit $a\to 0$, we recover the commutative co-ordinate. Also, ${\hat x}_\mu= 0$, 
when $x_\mu=0$. But in the non-commutative space-time, one cannot localise a particle below the length scale set by the 
non-commutativity parameter and thus when $a\ne 0$, taking the limit $x_\mu\to0$ is not valid.

The $\kappa$-deformed metric, $\hat{g}_{\mu\nu}$ is constructed by defining generalised commutation relations between deformed the phase space co-ordinates as
\begin{equation}\label{a1}
[\hat{x}_{\mu},\hat{P}_{\nu}]=i\hslash\hat{g}_{\mu\nu}.
\end{equation}
Here $\hat{P}_{\mu}$ is realised as $\hat{P}_{\mu}=g_{\alpha\beta}(\hat{y})p^{\alpha}\varphi_{\mu}^{\beta}(p)$, where $\hat{y}$ is an auxilary $\kappa$-deformed space-time co-ordinate satisfying $[\hat{y}_{\nu},\hat{x}_{\mu}]=0$ and hence any function of $\hat{y}_{\mu}$ also commutes with $\hat{x}_{\mu}$, i.e, $[f(\hat{y}),\hat{x}_{\mu}]=0$ \cite{tajron}. Thus using Eq.(\ref{real}) and Eq.(\ref{soln}), we find $\hat{y}_{\mu}$ and $f(\hat{y})$ as
\begin{equation}\label{a2}
\hat{y}_{\mu}=x_{\mu}+\alpha\frac{x\cdot pa_{\mu}}{\hslash}+\beta \frac{x\cdot ap_{\mu}}{\hslash}+(\alpha+1) \frac{x_{\mu}a\cdot p}{\hslash},
\end{equation}
\begin{equation}\label{a3}
f(\hat{y})=f(x)+\alpha\Big(\frac{a}{\hslash}\cdot\frac{\partial f}{\partial x}\Big)(x\cdot p)+\beta\Big(\frac{a}{\hslash}\cdot x\Big)\Big(\frac{\partial f}{\partial x}\cdot p\Big)+(\alpha+1)\Big(x\cdot\frac{\partial f}{\partial x}\Big)\Big(\frac{a}{\hslash}\cdot p\Big),
\end{equation}
respectively. Now we expand the LHS of Eq.(\ref{a1}) using the above defined realisation of 
$\hat{x}_{\mu}$ and $\hat{P}_{\nu}$, and we obtain
\begin{equation}\label{a4}
\hat{g}_{\mu\nu}=g_{\alpha\beta}(\hat{y})\Big(p^{\beta}\frac{\partial \varphi^{\alpha}_{\nu}}{\partial p^{\sigma}}\varphi_{\mu}^{\sigma}+\varphi_{\mu}^{\alpha}\varphi_{\nu}^{\beta}\Big). 
\end{equation}
Using Eq.(\ref{a3}) we re-express $g_{\mu\nu}(\hat{y})$ in terms of $x$ as
\begin{equation}\label{a5}
g_{\mu\nu}(\hat{y})=g_{\mu\nu}(x)+\alpha\Big(\frac{a}{\hslash}\cdot\frac{\partial g_{\mu\nu}}{\partial x}\Big)(x\cdot p)+\beta\Big(\frac{a}{\hslash}\cdot x\Big)\Big(\frac{\partial g_{\mu\nu}}{\partial x}\cdot p\Big)+(\alpha+1)\Big(x\cdot\frac{\partial g_{\mu\nu}}{\partial x}\Big)\Big(\frac{a}{\hslash}\cdot p\Big).
\end{equation}
Substituting Eq.(\ref{soln}) and Eq.(\ref{a5}) in Eq.(\ref{a4}), we get the expression for the $\kappa$-deformed metric, valid up to first order in $a$, as
\begin{equation}\label{a6}
\begin{split}
\hat{g}_{\mu\nu}&=g_{\mu\nu}+\alpha\bigg(p^{\beta}g_{\nu\beta}\frac{a_{\mu}}{\hslash}+2g_{\mu\nu}\Big(\frac{a}
{\hslash}\cdot p\Big)+\Big(\frac{a}{\hslash}\cdot\frac{g_{\mu\nu}}{\partial x}\Big)x\cdot p\bigg)+\beta\bigg(p^{\beta}\frac{a^{\alpha}}{\hslash}g_{\alpha\beta}\eta_{\mu\nu}+\frac{a^{\beta}}{\hslash}p_{\nu}g_{\mu\beta}+\frac{a^{\beta}}{\hslash}p_{\mu}g_{\nu\beta}\\&+\Big(\frac{a}{\hslash}\cdot x\Big)\Big(\frac{\partial g_{\mu\nu}}{\partial x}\cdot p\Big)\bigg)
+(\alpha+1)\bigg(2g_{\mu\beta}p^{\beta}\frac{a_{\nu}}{\hslash}+p^{\alpha}\frac{a_{\mu}}{\hslash}g_{\alpha\nu}+\Big(x\cdot\frac{\partial g_{\mu\nu}}{\partial x}\Big)\Big(\frac{a}{\hslash}\cdot{p}\Big)\bigg).
\end{split}
\end{equation}
We notice that in the limit $a\to 0$ (also in the limit $\frac{a}{\hslash}\to 0$), we recover the commutative metric, i,e, $\hat{g}_{\mu\nu}$ becomes $g_{\mu\nu}$. Further we note that for $a\ne 0$, in the limit $\hbar\to 0$, the metric 
$\hat{g}_{\mu\nu}$ diverges.

We note here that the above obtained $\kappa$-deformed metric depends on the momentum $p_\mu$, apart 
from the space-time co-ordinates. In 
the earlier work on maximal acceleration in the $\kappa$-space-time \cite{vishnu4} (with a different realisation that depends only on 
the deformation parameter as well as metric deformation energy scale), the momentum dependency of the 8-dimensional 
deformed metric has been obtained by taking the direct sum of the deformed space-time metric and the deformed 
metric of the momentum space (obtained from $\kappa$-deformed dispersion relation). Here the realisation defined 
in Eq.(\ref{soln}) depends on the momenta and hence this induces momentum dependent terms in the 
$\kappa$-deformed metric (given in Eq.(\ref{a6})), though the 4-dimensional deformed metric is 
constructed differently compared to the one used in  \cite{vishnu4}, which was constructed in 8-dimensional space by taking 
direct sum of deformed 4-dimensional space-time metric and metric in the corresponding momentum space.

\section{$\kappa$-deformed maximal acceleration} 

In this section, we start with the general expression for the line-element in $\kappa$-deformed space-time and using this we then obtain the line-element for the $\kappa$-deformed Minkowski space-time, valid up to first order in $a$. Imposing the condition that the events in the 4-dimensional $\kappa$-deformed Minkowski space-time to be causally connected, we derive the explicit form of the maximal acceleration, emerging from the $\kappa$-deformed space-time geometry, valid up to first order in $a$.

The $\kappa$-deformed line element is defined in terms of the deformed metric and the differential of deformed space-time co-ordinate as 
\begin{equation}\label{a8}
d\hat{s}^2=\hat{\eta}_{\mu\nu}d\hat{x}^{\mu}d\hat{x}^{\nu}.
\end{equation}
We evaluate $d\hat{x}^{\mu}$ by taking the differential of Eq.(\ref{x}) and replacing $g_{\mu\nu}$ with $\eta_{\mu\nu}$ in Eq.(\ref{a6}) we find $\hat{\eta}_{\mu\nu}$. Using this $\hat{\eta}_{\mu\nu}$ and $d\hat{x}^{\mu}$ in Eq.(\ref{a8}) we obtain the $\kappa$-deformed Minkowskian line element, valid up to first order in $a$, as
\begin{equation}\label{a9}
\begin{split}
d\hat{s}^2&={\eta}_{\mu\nu}dx^{\mu}dx^{\nu}+\alpha\bigg(\Big(2p_{\nu}\frac{a_{\mu}}{\hslash}+4\eta_{\mu\nu}\frac{a}{\hslash}\frac{E}{c}+2p_{\mu}\frac{a_{\nu}}{\hslash}\Big)dx^{\mu}dx^{\nu}+\frac{a}{\hslash}\eta_{\mu\nu}\frac{dE}{c}dx^{\mu}x^{\nu}+(x\cdot dx)\frac{a}{\hslash}\frac{dE}{c}\\&+2\frac{a}{\hslash}cdt(dx\cdot p)+2(x\cdot dp)cdt\frac{a}{\hslash}\bigg)+\beta\bigg(\Big(\eta_{\mu\nu}\frac{a}{\hslash}\frac{E}{c}+\frac{a_{\mu}}{\hslash}p_{\nu}+\frac{a_{\nu}}{\hslash}p_{\mu}\Big)dx^{\mu}dx^{\nu}+2\frac{a}{\hslash}(dx\cdot p)cdt\\&+2\frac{a}{\hslash}(dx\cdot dp)ct\bigg)+\bigg(\frac{3}{2}\Big(p_{\mu}\frac{a_{\nu}}{\hslash}+p_{\nu}\frac{a_{\mu}}{\hslash}\Big)dx^{\mu}dx^{\nu}+2\frac{a}{\hslash}cdt(dx\cdot p)+2\frac{a}{\hslash}cdt(x\cdot dp)\bigg).
\end{split}
\end{equation}
Note that the deformed line element depends on $E$ and $\vec{p}$, which enter through the realisation of non-comutative coordiantes(see Eqn.(\ref{x})). For the remaining analysis, without loss of generality, we work with $1+1$ dimenisonal $\kappa$-deformed space-time. The deformed line element in $1+1$ dimension is
\begin{equation}\label{a10}
\begin{split}
d\hat{s}^2=&c^2dt^2-dz^2+\alpha\bigg(6\frac{a}{\hslash}cdt\Big(Edt-pdz\Big)+4\frac{a}{\hslash}\frac{E}{c}\Big(c^2dt^2-dz^2\Big)+2\frac{a}{\hslash}\frac{dE}{c}\Big(c^2tdt-zdz\Big)\\&+2\frac{a}{\hslash}cdt\Big(tdE-zdp\Big)\bigg)+\beta\bigg(\frac{a}{\hslash}\frac{E}{c}\Big(c^2dt^2-dz^2\Big)+4\frac{a}{\hslash}cdt\Big(Edt-pdz\Big)+2\frac{a}{\hslash}ct\Big(dEdt-dpdz\Big)\bigg)\\&+5\frac{a}{\hslash}cdt\Big(Edt-pdz\Big)+2\frac{a}{\hslash}cdt\Big(tdE-zdp\Big).
\end{split}
\end{equation}
Now let us consider a time-like event, so that $d\hat{s}^2\geq 0$ and we divide the above equation by $dt^2$, denote $\frac{dz}{dt}$ as $v$, which is the velocity and $\frac{dp}{dt}=\frac{mA}{(1-v^2/c^2)^{3/2}}$, where $m$ is the rest mass of the particle and $A$ is the proper acceleration, respectively. Using these definitions, above expression 
takes the form
\begin{equation}\label{a11}
\begin{split}
&c^2-v^2+\alpha\bigg(6\frac{a}{\hslash}c\Big(E-pv\Big)+4\frac{a}{\hslash}\frac{E}{c}\Big(c^2-v^2\Big)+2\frac{a}{\hslash}\frac{dE}{dt}\frac{1}{c}\Big(c^2t-zv\Big)+2\frac{a}{\hslash}c\Big(t\frac{dE}{dt}-z\frac{mA}{(1-v^2/c^2)^{3/2}}\Big)\bigg)
+\\&\beta\bigg(\frac{a}{\hslash}\frac{E}{c}\Big(c^2-v^2\Big)+4\frac{a}{\hslash}c\Big(E-pv\Big)
+2\frac{a}{\hslash}ct\Big(\frac{dE}{dt}-v\frac{mA}{(1-v^2/c^2)^{3/2}}\Big)\bigg)
+\bigg(5\frac{a}{\hslash}c\Big(E-pv\Big)+\\
&2\frac{a}{\hslash}c\Big(t\frac{dE}{dt}-z\frac{mA}{(1-v^2/c^2)^{3/2}}\Big)\bigg)\geq 0
\end{split}
\end{equation}
Note that three terms in the above equation depend on the acceleration $A$. Of these, two terms are of the form $zA$ and the third is of the form $vA$. We can trace these terms to the $a$ dependent corrections involving $x\cdot dx=|{\vec x}||{\vec {dx}}|cos\theta$ and $dx\cdot dp= |{\vec {dx}}||{\vec {dp}}|cos\theta$ terms, respectively in Eqn.(\ref{a9}). To calculate the maximum acceleration, we take $cos\theta=1$ and to avoid confusion, 
explicitly write $|z|$ in place of $z$, in equations below.

%$\clubsuit\clubsuit$Note that there terms in the above equatons which depend on the acceleration $A$. Of these, 
%some terms are of the form $pv,~zv,~zA$ and $vA$. We can trace these terms to the $a$ depended corrections 
%involving $p\cdot dx= |{\vec {p}}||{\vec {dx}}|cos\theta,
%~x\cdot dx=|{\vec x}||{\vec {dx}}|cos\theta,~x\cdot dp=|{\vec x}||{\vec {dp}}|cos\theta$ 
%and $dx\cdot dp= |{\vec {dx}}||{\vec {dp}}|cos\theta$ terms, respectively in Eqn.(\ref{a9}). To calculate the 
%maximum acceleration, we take $cos\theta=1$ and to avoid confusion, 
%explicitly write $|z|$ in place of $z$, in equatons below.$\clubsuit\clubsuit$

We divide the Eq.(\ref{a11}) throughout by $c^2-v^2$ and use the commutative dispersion relation to re-write $\frac{dE}{dt}$ as $\frac{dE}{dt}=\frac{pc^2}{E}\frac{dp}{dt}=\frac{pc^2}{E}\frac{mA}{(1-v^2/c^2)^{3/2}}$. Note that since we are working to the order $a$ in our analysis, we need to use only commutative expression for $\frac{dE}{dt}$ as $\frac{dE}{dt}$ terms in Eq.(\ref{a11}) are already of order $a$. Thus we re-express the above Eq.(\ref{a11}) as
\begin{equation}\label{a12}
\begin{split}
&1+\alpha\bigg(6\frac{a}{\hslash}\frac{c(E-pv)}{c^2-v^2}+4\frac{a}{\hslash}\frac{E}{c}
+2\frac{a}{\hslash}\frac{pc}{E}\frac{mA}{(1-v^2/c^2)^{3/2}}\frac{(c^2t-|z|v)}{c^2-v^2}
+2\frac{a}{\hslash}\frac{c}{c^2-v^2}\Big(t\frac{pc^2}{E}-|z|\Big)\frac{mA}{(1-v^2/c^2)^{3/2}}\bigg)+\\&
\beta\bigg(\frac{a}{\hslash}\frac{E}{c}+4\frac{a}{\hslash}\frac{c(E-pv)}{c^2-v^2}
+2\frac{a}{\hslash}\frac{ct}{c^2-v^2}\Big(\frac{pc^2}{E}-v\Big)\frac{mA}{(1-v^2/c^2)^{3/2}}\bigg)
+\\&\bigg(5\frac{a}{\hslash}\frac{(E-pv)}{c^2-v^2}
+2\frac{a}{\hslash}\frac{c}{c^2-v^2}\Big(t\frac{pc^2}{E}-|z|\Big)\frac{mA}{(1-v^2/c^2)^{3/2}}\bigg)
\geq 0,
\end{split}
\end{equation}
which is valid up to first order in $a$. As in the Minkowski space-time, maximum allowed velocity for any particle in the $\kappa$-deformed space-time is $c$. Hence acceleration of a particle can reach the maximum value, if it starts with zero velocity. Thus, to calculate the maximum acceleration, we choose instantaneous rest frame of the particle in which $v=0$ \cite{cai1,vishnu4}. Note that setting $v=0$ also set $p=mv=0$ and many terms in the above equation vanish, in particular, all 
$t$ dependent terms in the above equation vanish. With these, from Eq.(\ref{a12}) we derive the expression for magnitude of the maximal acceleration $\mathcal{A}$, valid up to first order in $a$, as
\begin{equation}\label{a13}
\mathcal{A}\leq \frac{c}{2(\frac{a}{\hslash})|z|m}\frac{1}{(1+\alpha)}\bigg[1+5\frac{a}{\hslash}\frac{E}{c}\big(1+\beta+2\alpha\big)\bigg].
\end{equation} 
Here $E$ represents the back ground energy scale of non-commutativity which entered through the deformed metric. We note that one of the terms above depends on $\hbar$ while the other is 
independent of $\hbar$. The second term depends on the non-commutative parameters, $\alpha$ and $\beta$ and thus arise purely due to the non-commutativity of the underlying space-time, which is a new feature compared to the results obtained from $8$-dimensional line element, both in the commutative space-time \cite{cai1} as well as in the $\kappa$-space-time \cite{vishnu4}. Thus we find, with $a\ne 0$, limit $\hbar\to 0$, ${\mathcal A}\le\frac{5~E(1+\beta+2\alpha)}{2|z|m(1+\alpha)}$, which should be contrasted with the results obtained in\cite{vishnu4} as well as in\cite{cai1}.

Note that $\mathcal{A}$ being the magnitude of maximal acceleration, it is a positive quantity. Thus we see that for $\alpha>-1$, we have $~\beta>-(1+2\alpha+\frac{c\hbar}{5aE})$ and for $\alpha<-1$, we will have $~\beta<-(1+2\alpha+\frac{c\hbar}{5aE})$. Note that in the limit $a\to 0$, $\mathcal{A}\to\infty$ as expected in the commutative limit. The appearence of $|z|$ term in $\mathcal{A}$ is due to the presence of $x_{\mu}$ in the realisation, i.e Eq.(\ref{x}) ($E$ can also be traced back to $p_0$ appearing in Eq.(\ref{x}), which defines deformed co-ordinates.). Thus we find the maximal acceleration induced by the non-commutativity depends inversely on the modulus of the spatial co-ordinate. Since, in non-commutative space-time, one 
cannot localise a particle below the length scale set by the non-commutative parameter, $|z|$ 
cannot be less than $a$, i.e., $|z|\ge a$. Thus taking the limit $z\to 0$ in the above with $a\ne 0$ is not allowed. Also, for a particle moving under the influence of a force exerted by another object, $|z|$ represents the minimum distance of separation between them. As stated above, with $a\ne 0$, in the limit $\hbar\to 0$, we find ${\mathcal A}\le\frac{5~E(1+\beta+2\alpha)}{2|z|m(1+\alpha)}$. The deformation energy $E$ appearing in this limit is of the order of Planck energy and thus  $\mathcal{A}$ will be very large unless $\frac{1+\beta+2\alpha}{1+\alpha}$ is very small. We observe that the limit $\hbar\to 0$, the bounds on $\alpha$ as well as $\beta$ become either $\alpha>-1,~\beta>-(1+2\alpha)$ or $\alpha<-1,~\beta<-(1+2\alpha)$. 

We see from the above equation that in the commutative limit (i.e., $a\to 0$), $\mathcal{A}\to\infty$ and with $a\ne 0$, the classical limit where $\hbar\to 0$ gives the bound ${\mathcal A}\le\frac{5~E(1+\beta+2\alpha)}{2|z|m(1+\alpha)}$. The numerical parameters, $\alpha$ and $\beta$ enter the Eq.(\ref{a13}) through the realisation of the non-commutative co-ordinate (see Eq.(\ref{x})) and have to be fixed by experiments or from other theoretical inputs. We note that the expression for the maximal acceleration in Eq.(\ref{a13}) in the $\kappa$-space-time is derived from the 4-dimensional line element unlike in \cite{cai1} and other earlier works where the maximal acceleration in commutative space-time was obtained from line element defined in $8$-dimensional phase-space. The maximum acceleration obtained in \cite{cai1}) depends on $\hbar$ and goes to infinity in the classical limit, i.e., when $\hbar\to 0$. These features should be contrasted with the maximal acceleration obtained here in Eq.(\ref{a13}) 
which is derived from the deformed 4-dimensional metric. Note that the maximal acceleration obtained in Eq.(\ref{a13}) above arises due to the geometry of the non-commutative space-time itself.

\section{$\kappa$-deformed geodesic equation}

In this section, we derive the $\kappa$-deformed geodesic equation and obtain its Newtonian limit. For this, we first define the $\kappa$-deformed Christoffel symbol by replacing the commutative metric with the $\kappa$-deformed metric in the definition of Christoffel symbol and then using this 
$\kappa$-deformed Christoffel symbol derive the geodesic equation of a test particle in the $\kappa$-deformed space-time, valid up to first order in $a$. It is to be noted that one can also obtain the $\kappa$-deformed geodesic equation in an alternate way by generalising the Feynman's approach to $\kappa$-deformed space-time. Using this approach, the curvature effects in $\kappa$-deformed space-time were studied and the geodesic equation was derived earlier in \cite{tajron}. Such non-vanishing curvature like contributions are also observed in Moyal space-time \cite{victor}.

We analyse the special case of Eq.(\ref{a6}) by setting $\beta=0$ (see appendix.A for the justification) and hence the expression for the deformed metric becomes
\begin{equation}\label{c1}
\begin{split}
\hat{g}_{\mu\nu}&=g_{\mu\nu}+\alpha\bigg(p^{\beta}g_{\nu\beta}\frac{a_{\mu}}{\hslash}
+2g_{\mu\nu}\Big(\frac{a}{\hslash}\cdot p\Big)+\Big(\frac{a}{\hslash}\cdot\frac{g_{\mu\nu}}{\partial x}\Big)x\cdot p\bigg)
+(\alpha+1)\bigg(2g_{\mu\beta}p^{\beta}\frac{a_{\nu}}{\hslash}+p^{\alpha}\frac{a_{\mu}}{\hslash}g_{\alpha\nu}+\Big(x\cdot\frac{\partial g_{\mu\nu}}{\partial x}\Big)\Big(\frac{a}{\hslash}\cdot{p}\Big)\bigg).
\end{split}
\end{equation}
The Christoffel symbol in $\kappa$-deformed space-time is defined as
\begin{equation}\label{c2}
\hat{\Gamma}_{\nu\lambda}^{\mu}=\frac{1}{2}\hat{g}^{\mu\rho}\big(\partial_{\nu}\hat{g}_{\rho\lambda}+\partial_{\lambda}\hat{g}_{\nu\rho}-\partial_{\rho}\hat{g}_{\nu\lambda}\big)
\end{equation}
Substituting the deformed metric given in Eq.(\ref{c1}) in the above equation, we calculate the deformed Christoffel symbol, valid up to first order in $a$, as
\begin{equation}\label{c3}
\hat{\Gamma}_{\nu\lambda}^{\mu}={\Gamma}_{\nu\lambda}^{\mu}+\frac{1}{2}\frac{a}{\hslash}
\Big(m_2\mathcal{B}^{\mu}_{\nu\lambda\sigma}\frac{dx^{\sigma}}{d\tau}+\frac{E}{c}\mathcal{C}^{\mu}_{\nu\lambda}\Big),
\end{equation}   
where,
\begin{equation}\label{c4}
\begin{split}
\mathcal{B}^{\mu}_{\nu\lambda\sigma}&=\alpha g^{\mu\rho}\bigg(2\partial_{\nu}g_{\sigma[\rho}\delta_{\lambda] 0}+2\partial_{\lambda}g_{\sigma[\rho}\delta_{\nu] 0}-2\partial_{\rho}g_{\sigma[\nu}\delta_{\lambda]0}+\frac{1}{c}\partial_{\nu}\Big(\frac{\partial g_{\rho\lambda}}{\partial t}x_{\sigma}\Big)+\frac{1}{c}\partial_{\lambda}\Big(\frac{\partial g_{\nu\rho}}{\partial t}x_{\sigma}\Big)-\frac{1}{c}\partial_{\rho}\Big(\frac{\partial g_{\nu\lambda}}{\partial t}x_{\sigma}\Big)\bigg)\\
&+\frac{3}{2}g^{\mu\rho}\bigg(\partial_{\nu}g_{\sigma[\rho}\delta_{\lambda]0}+\partial_{\lambda}g_{\sigma[\rho}\delta_{\nu]0}-\partial_{\rho}g_{\sigma[\nu}\delta_{\lambda]0}\bigg)+\frac{3}{2}\Big(\partial_{\nu}g_{\rho\lambda}+\partial_{\lambda}g_{\nu\rho}-\partial_{\rho}g_{\nu\lambda}\Big)g_{\sigma}^{[\mu}\delta^{\rho]0}\\
&+\alpha \Big(\partial_{\nu}g_{\rho\lambda}+\partial_{\lambda}g_{\nu\rho}-\partial_{\rho}g_{\nu\lambda}\Big)\Big(2\delta^{0[\rho}g^{\nu]\sigma}+\frac{1}{c}\frac{\partial g^{\mu\rho}}{\partial t}x^{\sigma}\Big),
\end{split}
\end{equation}
and
\begin{equation}\label{c5}
\begin{split}
\mathcal{C}^{\mu}_{\nu\lambda}=&\bigg(4\alpha g^{\mu\rho}+\big(\alpha+1\big)\Big(x\cdot \frac{\partial g^{\mu\rho}}{\partial x}\Big)\bigg)\Big(\partial_{\nu}g_{\rho\lambda}+\partial_{\lambda}g_{\nu\rho}-\partial_{\rho}g_{\nu\lambda}\Big)\\&+\big(\alpha+1\big)g^{\mu\rho}\bigg(\partial_{\nu}\Big(x\cdot \frac{\partial g_{\rho \lambda}}{\partial x}\Big)+\partial_{\lambda}\Big(x\cdot \frac{\partial g_{\nu \rho}}{\partial x}\Big)-\partial_{\rho}\Big(x\cdot \frac{\partial g_{\nu \lambda}}{\partial x}\Big)\bigg).
\end{split}
\end{equation}
Thus we see that the deformed Christoffel symbol, valid up to first order in $a$, contains two non-commutative correction terms, in which first one depends on the four-momentum of the test particle whereas the second correction term is dependent on $E$, the non-commutative energy scale. 

Using the geodesic equation in the $\kappa$-deformed space-time given by
\begin{equation}\label{c6}
\frac{d^2\hat{x}^{\mu}}{d\tau ^2}+\hat{\Gamma}^{\mu}_{\nu\lambda}\frac{d\hat{x}^{\nu}}{d\tau}\frac{d\hat{x}^{\lambda}}{d\tau}=0,
\end{equation} 
and $\frac{d\hat{x}^{\mu}}{d\tau},~\frac{d^2\hat{x}^{\mu}}{d\tau^2}$ calculated from Eq.(\ref{x}) and Eq.(\ref{c3}), the $\kappa$-deformed geodesic equation, valid up to first order in $a$ is derived and it is given by
\begin{equation}\label{c7}
\frac{d^2x^{\mu}}{d\tau ^2}+\Gamma^{\mu}_{\nu\lambda}\frac{dx^{\nu}}{d\tau}\frac{dx^{\lambda}}{d\tau}+\frac{a}{\hslash}\mathcal{D}^{\mu}_{\nu\lambda}\frac{dx^{\nu}}{d\tau}\frac{dx^{\lambda}}{d\tau}+\frac{a}{\hslash}\mathcal{E}^{\mu}_{\nu\lambda\sigma}\frac{dx^{\nu}}{d\tau}\frac{dx^{\lambda}}{d\tau}\frac{dx^{\sigma}}{d\tau}+\frac{a}{\hslash}\mathcal{F}^{\mu}_{\nu\lambda}\frac{d^2x^{\nu}}{d\tau^2}\frac{dx^{\lambda}}{d\tau}=0,
\end{equation} 
where,
\begin{equation}\label{c8}
\begin{split}
\mathcal{D}^{\mu}_{\nu\lambda}&=2\alpha\frac{E_1}{c}\Gamma^{\mu}_{\nu\lambda}+\frac{1}{2}\frac{E}{c}\mathcal{C}^{\mu}_{\nu\lambda},\\
\mathcal{E}^{\mu}_{\nu\lambda\sigma}&=m_1(\alpha+1)\Gamma^{\mu}_{\rho\gamma}\Big(\delta_{\sigma}^{\rho}\delta^{\gamma 0}+\delta_{\sigma}^{\gamma}\delta^{\rho 0}\Big)\eta_{\nu\lambda}+\frac{1}{2}m_2\mathcal{B}^{\mu}_{\nu\lambda\sigma},\\
\mathcal{F}^{\mu}_{\nu\lambda}&=3m_1\big(\alpha+1\big)\eta_{\nu\lambda}\delta^{\mu 0}+m_1\big(\alpha+1\big)x_{\nu}\Big(\delta_{\lambda}^{\rho}\delta^{\gamma 0}+\delta_{\lambda}^{\gamma}\delta^{\rho 0}\Big)\Gamma^{\mu}_{\rho\gamma}.
\end{split}
\end{equation}
Note that in the above definitions, $m_1,E_1$ and $m_2,E_2$ correspond to mass as well as energy coming from the first order non-commutative corrections of $\frac{d\hat{x}_{\mu}}{d\tau}$ and $\hat{g}_{\mu\nu}$ respectively.

\subsection{$\kappa$-deformed Newtonian limit}

In this subsection, we derive the $\kappa$-deformed Newton's force equation, valid up to first order in $a$, by using the same approach used in the commutative case. We obtain a constraint on the parameter $\alpha$ present in the maximal acceleration expression, by comparing the first order correction term in $\kappa$-deformed Newton's equation with the experimental bound on the violation of equivalence principle.

The $\kappa$-deformed geodesic equation, obtained in Eq.(\ref{c7}), is written in terms of commutative space-time co-ordinate and the deformation parameter, so we use the same approach as that in the commutative case to obtain the $\kappa$-deformed Newtonian limit of Eq.(\ref{c7}). In this approach, one assumes
\begin{itemize}
\item the test particles are moving slowly, i.e, $\frac{dx_i}{d\tau}<<\frac{dx_0}{d\tau}$.
\item the metric is static, i.e, $\frac{\partial g_{\mu\nu}}{\partial t}=0$.
\item gravitational field is weak and the metric is linearised as 
$g_{\mu\nu}=\eta_{\mu\nu}+h_{\mu\nu}$ where $|h_{\mu\nu}|<<1$.
\end{itemize}
Using these conditions in Eq.(\ref{c7}), we get
\begin{equation}\label{c10}
\frac{d^2x^0}{d\tau^2}\left(1+3\frac{a}{\hslash}m_1(\alpha+1)\frac{dx^0}{d\tau}\right)=0.
\end{equation}
and
\begin{equation}\label{c10a}
\begin{split}
\frac{d^2x^j}{d\tau^2}\left(\delta^{ij}+\frac{a}{\hslash}m_1(\alpha+1)x_j\partial_ih_{00}\frac{dx^0}{d\tau}\right)&=-\frac{1}{2}\partial_ih_{00}\frac{dx^0}{d\tau}\frac{dx^0}{d\tau}
\\&-\frac{1}{2}\frac{a}{\hslash}\bigg(\alpha\Big(2\frac{E_1}{c}+4\frac{E}{c}\Big)\partial_ih_{00}+(\alpha+1)\frac{E}{c}\partial_i\Big(x\cdot\frac{\partial h_{00}}{\partial x}\Big)\bigg)\frac{dx^0}{d\tau}\frac{dx^0}{d\tau}
\\&-\frac{1}{2}\frac{a}{\hslash}\Big(2m_1\left(\alpha+1\right)+m_2\left(4\alpha+3\right)\Big)\partial_ih_{00}\frac{dx^0}{d\tau}\frac{dx^0}{d\tau}\frac{dx^0}{d\tau}
\\&-\frac{a}{\hslash}m_1\left(\alpha+1\right)x_0\partial_ih_{00}\frac{d^2x^0}{d\tau^2}\frac{dx^0}{d\tau}.
\end{split}
\end{equation}
From the Eq.(\ref{c10}), we find that the terms inside the bracket cannot be zero as it lead to inconsistency in the commutative limit and therefore we infer that
\begin{equation}\label{c10b}
\frac{d^2x^0}{d\tau^2}=0.
\end{equation}
By using the Eq.(\ref{c10b}) in Eq.(\ref{c10a}), we find that the last term on RHS vanishes. Further we multiply Eq.(\ref{c10a}) throughout with 
$\left(\delta^{ij}-\frac{a}{\hslash}m_1(\alpha+1)x_j\partial_ih_{00}\frac{dx^0}{d\tau}\right)$ and keep the terms linear in $a$ and $h_{00}$, and find
\begin{equation}\label{c10c}
\begin{split}
\frac{d^2x^i}{d\tau^2}&=-\frac{1}{2}\partial_ih_{00}\frac{dx^0}{d\tau}\frac{dx^0}{d\tau}-\frac{1}{2}\frac{a}{\hslash}\bigg(\alpha\Big(2\frac{E_1}{c}+4\frac{E}{c}\Big)\partial_ih_{00}+(\alpha+1)\frac{E}{c}\partial_i\Big(x\cdot\frac{\partial h_{00}}{\partial x}\Big)\bigg)\frac{dx^0}{d\tau}\frac{dx^0}{d\tau}\\&-\frac{1}{2}\frac{a}{\hslash}\Big(2m_1\left(\alpha+1\right)+m_2\left(4\alpha+3\right)\Big)\partial_ih_{00}\frac{dx^0}{d\tau}\frac{dx^0}{d\tau}\frac{dx^0}{d\tau}.
\end{split}
\end{equation}
This equation is reparametrisation invariant up to $\mathcal{O}(h)$, as in the case of the Newtonian limit of $\kappa$-deformed geodesic equation obtained in\cite{tajron}. In calculating the deformed Newtonian limit, we have already assumed that the test particle is moving slowly, hence we can take $E_1= m_1c^2$. Now using the Newtonian potential in the commutative space-time, i.e., $h_{00}=-\frac{2GM}{c^2r}$, we obtain the $\kappa$-deformed Newton's force equation, valid up to first order in $a$, as
\begin{equation}\label{c11}
\hat{F}^i=F^i\bigg(1+\frac{a}{\hslash}\left(\frac{E}{c}\big(3\alpha-1\big)+m_2c\big(4\alpha+3\big)+m_1c\big(4\alpha+2\big)\right)\bigg),
\end{equation}
where $F^i=-\frac{m_1MG}{r^2}$. Here we find that the Newton's force equation, defined in the $\kappa$-deformed space-time, valid up to first order in $a$, picks up two correction terms, of which one depends on the $\kappa$-deformed background energy scale and the other one depends on the rest mass of the particle. Note that the $\kappa$-deformed force equation has only radial component as in the commutative case.

The realisation given in Eq.(\ref{soln}) itself has the dependency on mass and deformation energy (unlike the $\varphi=e^{-aE}$ realisation, where the dependency is on the deformation energy only). Thus the deformed force equation obtained from this realisation has also two non-commutative correction terms, depending on mass of particle and deformation energy. On the other hand the deformed Newton's force equation, obtained in \cite{vishnu4}, (corresponding to $\varphi=e^{-aE}$ realisation) has only one non-commutative correction that depends on deformation energy.

We notice that the mass dependent correction term in the deformed Newton's force equation violates the equivalence principle and this change in ratio of gravitational to inertial mass is $\frac{am_1c(4\alpha+2)}{\hbar}$. Experimental limit on the violation of equivalence principle\cite{vep} is known with an accuracy of $10^{-13}$. Using this, we find $\frac{am_1c(4\alpha+2)}{\hbar}<10^{-13}$. Hence we obtain the bound as $\alpha<-0.5$, for a test particle of unit mass and for the deformation parameter ranging from $a=10^{-49}m$ to $a=10^{-21}m$. Thus, using this bound 
and the conditions $\alpha>-1$ and $\beta>-(1+2\alpha+\frac{c\hbar}{5aE})$ we get a strong bound on $\alpha$ as $-1<\alpha<-0.5$ for $\beta=0$.

\section{Conclusions}

In this paper, we have shown that in the $\kappa$-deformed space-time, for every massive particle, an upper cut-off value for the allowed acceleration exists. This is obtained by constructing the $4$-dimensional line element in the $\kappa$-deformed Minkowski space-time, valid up to first order in the deformation parameter, $a$ and analysing the condition for a causal connection between any two events. This should be contrasted with earlier works where maximal acceleration is derived from an $8$-dimensional line element, constructed by taking direct sum of the metric of the space-time and the metric of the momentum space. The maximal acceleration we obtained has two terms; one depends on the non-commutative parameters $a$ and $\alpha$, $\hbar$, the mass of the particle, a minimum distance of approach, while the other is independent of $\hbar$ but depends on the non-commutative parameter $\beta$ and the non-commutative energy scale, in addition to $\alpha$, the mass of the particle and minimum distance of approach. This second term, which is independent of $\hbar$ is purely due to non-commutativity of the space-time. This $\hbar$ independent correction is absent in the maximal acceleration obtained using the 8-dimensional line element of the $\kappa$-deformed space-time derived in \cite{vishnu4} and/or the one obtained in commutative space-time\cite{cai1}. This is a novel feature of our result.  We have seen that the maximal acceleration blows up as the non-commutative parameter $a$ goes to zero. But unlike the earlier results in the commutative space-time, in the limit, $\hbar\to 0$, the maximal acceleration goes to a finite value (when $a\ne 0$). This limiting value depends on the mass of the particle, distance of shortest approach and non-commutative energy scale $E$ and the non-commutative parameters $\alpha$ and $\beta$. Note that this classical limit (i.e., $\hbar\to 0$) does not exist for the maximal acceleration obtained in the commutative space-time or for the one obtained in \cite{vishnu4}. In\cite{toller}, the implication of the existence of maximum allowed acceleration was studied using the approach of the jet bundle, which also avoids the use of 8-dimensional line element. In this case, the usual Minkowski space-time is obtained in the limit where the maximum acceleration diverges to infinity as in the present 
case.

Note that the Unruh temperature $T=\frac{\hbar A}{2\pi kc},$ where $k$ is the Boltzmann constant, is shown to be unaffected by the $\kappa$-deformation of space-time \cite{vishnu1}. Using the value of $\mathcal{A}$ obtained in Eq.(\ref{a13}) in the expression for Unruh tempertaure, we see that the maximum temperature gets a non-commutative correction, which depends on the deformation parameter $a$ and back ground energy scale $E$. After substituting the reduced Compton wave length of the particle of mass $m$, i.e ${\lambda}_C=\frac{\hbar}{mc}$ for $2|z|$, we find
\begin{equation}
T_{max}=\frac{\hslash c}{2\pi ka}\frac{1}{(1+\alpha)}\bigg(1+5\frac{aE}{\hbar c}(1+\beta+2\alpha)\bigg).
\end{equation}
In the commutative limit, $a\to 0$, we find that $T_{max}\to \infty$ and in the limit, $\hbar \to 0$, $T_{max}\to\frac{5E(1+\beta+2\alpha)}{2\pi k(1+\alpha)}$, where the non-commutative parameters $\alpha$ as well as $\beta$ satisfy either $\alpha>-1,~\beta>-(1+2\alpha+\frac{c\hbar}{aE})$ or 
$\alpha<-1,~\beta<-(1+2\alpha+\frac{c\hbar}{aE})$.

In contrast to the result obtained here, we note that the maximal acceleration derived using an $8$-dimensional line element of the $\kappa$-deformed space \cite{vishnu4} reduces to a finite value obtained in \cite{cai1} when $a\to 0$ (i.e., in the commutative limit). Hence the corresponding maximal temperature will also reduce to the one obtained in \cite{sakharov, landi}. Thus we see a clear difference in the commutative limit of maximal acceleration and maximal temperature obtained here from deformed line element in $4$-dimensions and the corresponding limit of the $\kappa$-deformed maximal acceleration derived from $8$-dimensional line element obtained by generalising the approach of \cite{cai1} to $\kappa$-deformed space-time in\cite{vishnu4}.

\section{Appendix A}
In this appendix, we first obtain the $\kappa$-deformed geodesic equation for the case $\beta\neq 0$ and then analyse its Newtonian limit in the $\kappa$-deformed space-time, valid up to first order in $a$. We find that the deformed geodesic equation leads to consistent Newtonian limit only for $\beta=0$, justifying the calculations in section.4, where we have taken $\beta=0$. Here too the equation takes the same form as that obtained in 
Eq.(\ref{c7}), but now the quantities $\mathcal{B}^{\mu}_{\nu\lambda},~\mathcal{E}^{\mu}_{\nu\lambda\sigma},~\mathcal{F}^{\mu}_{\nu\lambda}$ pick up $\beta$ dependent correction terms and they are as follow   
\renewcommand{\thesection}{A}
\begin{equation}\label{c20}
\begin{split}
\mathcal{B}^{\mu}_{\nu\lambda\sigma}&=\alpha g^{\mu\rho}\bigg(2\partial_{\nu}g_{\sigma[\rho}\delta_{\lambda] 0}+2\partial_{\lambda}g_{\sigma[\rho}\delta_{\nu] 0}-2\partial_{\rho}g_{\sigma[\nu}\delta_{\lambda]0}+\frac{1}{c}\partial_{\nu}\Big(\frac{\partial g_{\rho\lambda}}{\partial t}x_{\sigma}\Big)+\frac{1}{c}\partial_{\lambda}\Big(\frac{\partial g_{\nu\rho}}{\partial t}x_{\sigma}\Big)-\frac{1}{c}\partial_{\rho}\Big(\frac{\partial g_{\nu\lambda}}{\partial t}x_{\sigma}\Big)\bigg)\\
&+\frac{3}{2}g^{\mu\rho}\bigg(\partial_{\nu}g_{\sigma[\rho}\delta_{\lambda]0}+\partial_{\lambda}g_{\sigma[\rho}\delta_{\nu]0}-\partial_{\rho}g_{\sigma[\nu}\delta_{\lambda]0}\bigg)+\frac{3}{2}\Big(\partial_{\nu}g_{\rho\lambda}+\partial_{\lambda}g_{\nu\rho}-\partial_{\rho}g_{\nu\lambda}\Big)g_{\sigma}^{[\mu}\delta^{\rho]0}\\
&+\alpha \Big(\partial_{\nu}g_{\rho\lambda}+\partial_{\lambda}g_{\nu\rho}-\partial_{\rho}g_{\nu\lambda}\Big)\Big(2\delta^{0[\rho}g^{\nu]\sigma}+\frac{1}{c}\frac{\partial g^{\mu\rho}}{\partial t}x^{\sigma}\Big)+ 
\beta g^{\mu\rho}\Big(\eta_{\rho[\lambda}\partial_{\nu]}g_{0\sigma}-\eta_{\nu\lambda}\partial_{\rho}g_{0\sigma}+\delta_{\sigma[\lambda}\partial_{\nu]}g_{\rho 0}\\
&-\partial_{\rho}g_{0[\lambda}\delta_{\nu]\sigma}+\delta_{\rho0}\partial_{[\nu}g_{\lambda]0}+\partial_{\sigma}g_{\rho[\lambda}\delta{\nu]0}
-\delta_{\rho 0}\partial_{\sigma}g_{\nu\lambda}+x_0\partial_{\rho}\partial_{[\nu}g_{\lambda]\rho}-x_0\partial_{\rho}\partial_{\sigma}g_{\nu\lambda}\Big)\\
&+\beta\Big(g_{0\sigma}\eta^{\mu\rho}+\delta_{\sigma}^{[\rho}g^{\mu]}_0+x_0\partial_{\sigma}g^{\mu\rho}\Big)\Big(\partial_{\nu}g_{\rho\lambda}+\partial_{\lambda}g_{\nu\rho}-\partial_{\rho}g_{\nu\lambda}\Big),\\
\mathcal{E}^{\mu}_{\nu\lambda\sigma}&=m_1(\alpha+1)\Gamma^{\mu}_{\rho\gamma}\Big(\delta_{\sigma}^{\rho}\delta^{\gamma 0}+\delta_{\sigma}^{\gamma}\delta^{\rho 0}\Big)\eta_{\nu\lambda}+\frac{1}{2}m_2\mathcal{B}^{\mu}_{\nu\lambda\sigma}+m_1\beta \delta_{\sigma 0}\Gamma^{\mu}_{\nu\lambda},\\
\mathcal{F}^{\mu}_{\nu\lambda}&=3m_1\big(\alpha+1\big)\eta_{\nu\lambda}\delta^{\mu 0}+m_1\big(\alpha+1\big)x_{\nu}\Big(\delta_{\lambda}^{\rho}\delta^{\gamma 0}+\delta_{\lambda}^{\gamma}\delta^{\rho 0}\Big)\Gamma^{\mu}_{\rho\gamma}+\beta m_1\Big(\delta_{\lambda 0}\delta^{\mu}_{\nu}+\delta_{\nu o}\delta^{\mu}_{\lambda}\Big)+\beta m_1x_0\Gamma^{\mu}_{\nu\lambda}.
\end{split}
\end{equation}
It is to be noted that $\mathcal{C}^{\mu}_{\nu\lambda}$ and $\mathcal{D}^{\mu}_{\nu\lambda}$ do not pick any $\beta$ dependent correction terms up to first order in $a$. Now we obtain the Newtonian limit for the geodesic equation, defined in the $\kappa$-deformed space-time, when $\beta\neq 0$ following the same method used in sec.(4.1). Thus Eq.(\ref{c10b}) becomes
\begin{equation}\label{c21}
 \frac{d^2x^0}{d\tau^2}-\frac{1}{2}\frac{a}{\hbar}\beta m_1x_0\partial_ih_{00}\frac{d^2x^i}{d\tau^2}\frac{dx^0}{d\tau}=0.
\end{equation}
Now we substitute $\frac{1}{c}\frac{dx^0}{d\tau}=1-\frac{1}{2}h_{00}$ in above equation, i.e., in Eq.(\ref{c21}), and we get an inconsistent result, i.e., 
$\frac{1}{2}\frac{a}{\hbar}\beta m_1x_0\partial_ih_{00}\frac{d^2x^i}{d\tau^2}\Big(1-\frac{1}{2}h_{00}\Big)=0$. 
For an accelerated particle, $\frac{d^2x^i}{d\tau^2}\ne 0$ and hence this can be satisfied only if $\partial_ih_{00}=0$ or $h_{00}=2$. But either of these conditions are inconsistent with the Newtonian limit (in the commutative space-time). This leads to the condition $\beta=0$. Similar inconsistencies, for the case $\beta\neq 0$, were also reported in \cite{zuhair}.

\renewcommand{\thesection}{7}
\section{Appendix B}
\renewcommand{\thesection}{B} 
In this section we derive the $\kappa$-deformed corrections to maximal acceleration, valid up to first order in $a$, using the $\kappa$-deformed uncertainity relation between position and momenta. In the commutative space-time, this method has been used \cite{heis} and  shown to re-produce the maximal acceleration obtained using $8$-dimensional line element in \cite{cai1}. Thus we expect this result to be different from the one obtained in Eq.(\ref{a13}), which is derived from a $4$-diemnsional line element of $\kappa$-space-time and has seen to have contribution due to non-commutativity alone.
  
The uncertainity relation between energy as well as velocity and that between energy as well as position are defined as \cite{heis},   
\begin{equation}\label{a}
  \Delta E\Delta v(t) \geq \frac{\hbar}{2}\frac{dv}{dt} 
\end{equation}
and
\begin{equation}\label{v}
 \Delta E\Delta x(t) \geq \frac{\hbar}{2}\frac{dx}{dt} 
\end{equation}
respectively. Multiplying the above uncertainity relations, given in Eq.(\ref{a}) and Eq.(\ref{v}), we get
\begin{equation}
 \big(\Delta E\big)^2\Delta x\frac{\Delta v}{v}\geq \frac{\hbar^2}{4}\mathcal{A}
\end{equation}
where we have used the definitions $v=\frac{dx}{dt}$ and $\mathcal{A}=\frac{dv}{dt}$. Now we express $\Delta E$ as $\Delta E=v\Delta p$, and we get
\begin{equation}
 \big(\Delta p\big)^2\Delta x\big({\Delta v}\big)v\geq \frac{\hbar^2}{4}\mathcal{A}.
\end{equation}
From the special theory of relativity we infer that the uncertainty in velocity of the particle cannot exceed the velocity of light, i.e, $(\Delta v)^2=<v^2>-<v>^2\leq v_{max}^2\leq c^2$, so we take $(\Delta v)v\leq c^2$. After using this relation in the above equation we find that 
\begin{equation}\label{un}
 \big(\Delta p\Delta x\big)^2c^2\geq \frac{\hbar^2}{4}\mathcal{A}\Delta x
\end{equation}
The $\kappa$-deformed uncertainty relation between $\Delta x$ and $\Delta p$ has been derived in \cite{tajron}, up 
to first order in $a$, as
\begin{equation}\label{ku}
 \Delta x\Delta p\geq \frac{\hbar}{2}\Big(1+\frac{amc}{\hbar}(2\alpha+\beta)\Big).
\end{equation}
Substituting Eq.(\ref{ku}) in Eq.(\ref{un}) and choosing $\Delta x=\lambda$, where $\lambda$ is the reduced Compton wavelength, we obtain expression for Caianiello's maximal acceleration in $\kappa$-deformed space-time, valid up to first order in $a$, as
\begin{equation}\label{A2}
 \mathcal{A}_{U}\leq \frac{mc^3}{\hslash}\Big(1+\frac{2amc}{{\hslash}}(2\alpha+\beta)\Big).
\end{equation}
We note that the above maximal acceleration is different from the one obtained using $4$-dimensional line element of the $\kappa$-deformed space-time in Eq.(\ref{a13}). Note that in the limt $a\to 0$, above ${\mathcal A}_{U}$ reduces to $\frac{mc^3}{\hslash}$, the value obtained in \cite{cai1} for the commutative case. Also in the limit $\hbar\to 0$, we find ${\mathcal A}_{U}\to\infty$, as in the commutative space-time. Thus the maximal acceleration obtained using uncertainity relation in Eq.(\ref{A2}) has the same commutative limit ($a\to 0$) as well as the classical limit ($\hbar\to 0$), as expected. This again clearly shows that the maximal acceleration obtained from the $\kappa$-deformed, $4$-dimensional line element in Eq.(\ref{a13}), has contribution which is purely due to the non-commutativity of the space-time and thus independent of $\hbar$.
 
\subsection*{\bf Acknowledgments}
EH thanks SERB, Govt. of India, for support through EMR/2015/000622. VR thanks Govt. of India, for support through DST-INSPIRE/IF170622.

\end{document}